# A remark on the use of Bell inequalities

J. F. Geurdes

C vd Lijnstraat 164 2593 NN, Den Haag, Netherlands.

han.geurdes@gmail.com

**Abstract.**

In the present paper it is demonstrated that Bell's expression for local hidden variable correlation allows one to derive the quantum correlation. This raises questions about the use of Bell inequalities in conclusions from experiments.

# Introduction

The early beginnings of quantum theory (QM) are marked by questions of interpretation. In a later stage, an important step was taken by Bell[1]. Based on Einstein's criticism of completeness[2], Bell formulated the following expression for the relation between distant spin measurements, $A_a$ and $A_b$.

$$C_{a,b} = -\int \rho(\lambda) A_a(\lambda) A_b(\lambda) d\lambda$$

Here, $\rho(\lambda) \geq 0$, is the distribution of the local hidden variable(s), $\lambda$ with, $\int \rho(\lambda) d\lambda = 1$. The measurement functions, $A_a(\lambda)$ and, $A_b(\lambda)$ in $\{-1,1\}$, are supposed to depend on the unitary measurement parameter vector, $a$, or, $b$ and the local hidden variable(s), $\lambda$. Bell demonstrated that the correlation obeys the following inequality,

$$|C_{a,b} - C_{a,c}| \leq 1 + C_{b,c}$$

which can be violated by the quantum correlation, $-(a.b) = -a_1 b_1 - a_2 b_2 - a_3 b_3$.

Despite of the simplicity of the derivation of the above inequality, in the present paper it is questioned whether all local hidden variable correlations in the form of the Bell correlation are unable to become equal to the quantum correlation, $-(a.b)$. If this proofs to be possible, the mathematical steps followed by Bell are not as unqualified universally valid as is widely supposed. The reader should note that the attempt of construction of counter-examples is a logically valid and widely accepted way to study the claim of generality of theorems. The present author has worked along this lines in previous reports, Geurdes [4].

# I. Model

In order to have a simple and straightforward probability model, let us take the negative exponential probability density which indisputably gives a classical probability density.

$$\rho_\Lambda(\lambda) = \begin{cases} e^{-\lambda}, & \lambda \in [0,\infty) \\ 0, & elsewhere \end{cases} \tag{1}$$

In addition, a Kronecker delta discrete probability density is defined.

$$\rho_M(\mu) = \delta_{\mu,r}, \tag{2}$$

with, $r \in \{-1,1\}$. This density gives either probability zero or unity to an 'event', $\mu \in \{-1,1\}$.

Hence, the quantum correlation for measurement functions, $A_a(\lambda,\mu)$, $A_b(\lambda,\mu)$ projecting in the set, $\{-1,1\}$ in terms of local hidden variables, is given by

$$C_{a,b} = -\sum_{\mu \in \{-1,1\}} \int_0^{+\infty} e^{-\lambda} A_a(\lambda,\mu) A_b(\lambda,\mu) d\lambda \tag{3}$$

Let us define, furthermore, $F_{a,b}(\lambda,\mu) = A_a(\lambda,\mu) A_b(\lambda,\mu)$ and note that the vectors $a = (a_1, a_2, a_3)$ and $b = (b_1, b_2, b_3)$ are unitary in the Euclidean metric, $|a| = |b| = 1$. This gives for, $F_{a,b}(\lambda,\mu) \in \{-1,1\}$, the following form of the correlation

$$C_{a,b} = -\sum_{\mu \in \{-1,1\}} \int_0^{+\infty} e^{-\lambda} F_{a,b}(\lambda,\mu) d\lambda \tag{4}$$

Let us restrict the argument to situations where $a_3 = b_3 = 0$. This is no substantial limitation to the generality of our conclusions. If, subsequently, it is noted that

$$1 - e^{-\lambda} = \int_0^\lambda e^{-x} dx$$

then, the correlation can be written as

$$C_{a,b} = -\sum_{\mu \in \{-1,1\}} \int_0^{+\infty} F_{a,b}(\lambda,\mu) \{\frac{\partial}{\partial \lambda}(1 - e^{-\lambda})\} d\lambda \tag{5}$$

Formally, partial integration then leads to

$$C_{a,b} = -[F_{a,b}(\lambda)(1-e^{-\lambda})]_0^{+\infty} + \int_0^{+\infty}(1-e^{-\lambda})\{\sum_{\mu\in\{-1,1\}} \frac{\partial}{\partial\lambda}F_{a,b}(\lambda,\mu)\}d\lambda \tag{6}$$

with, $F_{a,b}(\lambda) = \sum_{\mu\in\{-1,1\}} F_{a,b}(\lambda,\mu)\delta_{\mu,r}$. After some further evaluation, writing as a short-hand,

$$I_{a,b} = \int_0^{+\infty}(1-e^{-\lambda})\{\sum_{\mu\in\{-1,1\}} \frac{\partial}{\partial\lambda}F_{a,b}(\lambda,\mu)\}d\lambda,$$

we may obtain

$$C_{a,b} = -F_{a,b}(\infty) + I_{a,b} \tag{7}$$

Note that, $F_{a,b}(\infty) \in \{-1,1\}$. Suppose, to continue, that $I_{a,b} > 0$. This implies that, because, $|C_{a,b}| \leq 1$, we must have, $F_{a,b}(\infty) = 1$. In addition, if $I_{a,b} < 0$, then, because $|C_{a,b}| \leq 1$, $F_{a,b}(\infty) = -1$.

This implies that for a sign function, $sign(x) = 1, x \geq 0$ and $sign(x) = -1, x < 0$, equation (7) can be written as

$$C_{a,b} = sign(I_{a,b})(|I_{a,b}| - 1) \tag{8}$$

Let us subsequently try to apply the previous finding to an interesting physical situation. E.g. in the computer network model of Gill[1] [3], there is strict locality. This means that the A-wing only responds to A-wing information plus 'blind' source information and similar for the B-wing. Hence, there is no way that the source can carry information of the state of the B-wing to the A-wing and vice versa. This implies a strict separable response of the, $A_a(\lambda,\mu)$ from the, $A_b(\lambda,\mu)$ and vice versa.

The reader should note in this point of the argumentation that although information is strictly separated a convolution in the postulated hidden variable space is not forbidden. This convolution of ' information' is an operation on separately stored 'parts' that are blended together but not by the action of the source but in the postulated space of hidden variables. Let us postulate that the source only sends three variables, $S_0, S_1, S_2$, all three projecting in, $\{-1,1\}$, to both the A-wing and the B-wing and that they are part of the convolution. Furthermore, let us postulate a convolution operator, $\hat{\wp}$, as

$$\hat{\wp} = -\log\frac{1}{2}\sum_{S_1\in\{-1,1\}}\frac{1}{2}\sum_{S_2\in\{-1,1\}}\log^{-1} 2\log\log \tag{9}$$

---

[1] The author is aware of the vast literature on Bell's theorem and can only defend his selection of attention to Gill's model because of its completeness and easy to grasp transparency based on ideas of Bell.

If the reader feels some unease with the introduction of a convolution then he preferably may ask himself why this is physically impossible. If one rejects the existence of local hidden variables, then of course, one rejects convolution but if the only defense is 'I do not believe …therefore not true', that is religion in an authoritarian sense. Why cannot the source send (uniformly distributed) hidden variables to both A and B and thereby define a convolution. As long as there is no answer, ignoring the theoretical preferences of the critic at this point, then it is allowed.

The general reading of the operator shows that when the convolution operator works on a variable, $x$, $(x>0)$, then first the (general) logarithm is taken of the variable, $x$. After that, the result is introduced in a second logarithm and that result is multiplied by 2. Subsequently, the anti-logarithm, $\exp = \log^{-1}$ is taken, after which the result is summed over $S_1$, and, $S_2$ and divided by 4. Finally, the logarithm is taken over that result and a multiplication with, -1, is performed. This convolution operation is allowed for strictly separated experimental results that arose from the settings of the parameter vectors and the variables arising from the source. If we define

$$\zeta_A = a_1 S_1 S_0 + a_2 S_2$$
$$\zeta_B = b_1 S_1 + b_2 S_2 S_0$$
(10)

then, $\zeta_A$ and $\zeta_B$ comply to the idea of separated local conglomerates of local variables supplemented with $S_0, S_1, S_2$. If a reader wants to oppose to that, he or she is reminded to the fact that Bell's definition of measurement results, works in the same way. If it is contested that $\zeta_A$ and $\zeta_B$ are separate local entities, like the spin measurement functions, then how to uphold Bell's theorem?

Because the hidden variable space may 'work with' the source variables, $S_0, S_1, S_2$, and taking a logarithm or an 'exp' function cannot violate the locality, the following product of the convolution operator and the two exponentials

$$\hat{\wp} e^{\zeta_A} e^{\zeta_B}$$

cannot violate locality either. Given this configuration of parameters and operators, the locality of the measurements $A_a(\lambda, \mu)$, and, $A_b(\lambda, \mu)$ in the experiment will not be disturbed when $F_{a,b}(\lambda, \mu) = A_a(\lambda, \mu) A_b(\lambda, \mu)$ results into

$$F_{a,b}(\lambda) = S_0 \, sign(\lambda - \hat{\wp} e^{\zeta_A} e^{\zeta_B} - \log(4))$$
(11)

Although the following section is important, the reader is advised first to read section III and the subsequent discussion on locality vs non-locality and the violated instance of the CHSH inequality. The reason is that the violation of CHSH is based on the particular form of the local model without the necessity of having knowledge of the A and B measurement functions. I.e. there it is demonstrated that, based on purely its locality arguments, at least one critical violation is possible with the model.

# II Auxiliary specification

At this point in the argument, a reader may rightfully ask how to validate the form of the above equation starting from the measurement functions themselves. We have already established that numerically and physically the previous equation is allowed but then the question will arise which form the elementary functions $A_a(\lambda, \mu)$, and, $A_b(\lambda, \mu)$ must have.

In the past the author already has provided independent counter-models [4]. Of course they were met with criticism because of, in the author's view, overly confidence in Bell's manipulations of his expression of the correlation given in the previous introduction. Perhaps that those manipulations appear universally valid, but there is no independent proof of the fact that Bell's inequality and its 'derivatives', refer to *all* Local Hidden Variables. Why is it generally accepted impossible that not *all* but merely *most* of the LHV models fall under Bell's theorem, disallowing explanations as *simplicity* and *obviousness.*

In turn the author may claim that the analysis up until and including equation (11) plus [4] is sufficient evidence for the existence of LHV. The evidence given below is then simply an addition to [4]. In this sense it is redundant but can provide interesting insights. However, the reader is reminded that opposition to the formulas given below can be overcome by the models in [4] even though they do not 1-1 connect to Eq. (11). Moreover, the author draws for the moment the attention to a preprint [7] which shows that Hardy's paradox *can* be explained with the use of classical probability theory. Hence, a critical reader is also no longer safe to rely on, or hide behind, arguments outside the mathematics presented here and in [4] and [5].

Let us introduce the following symbol, $\langle 0 |$. This symbol refers to a hidden state that has a role similar to, $S_0, S_1, S_2$. Note that not by necessity this symbol is a quantum mechanical state although admittedly it looks like one. It can be a, n dimensional, row vector and $|0\rangle$ its transposed column. Let us assume furthermore that, $\langle 0|0\rangle = 1$, and that, $|0\rangle\langle 0| = \hat{1}$. Subsequently, the $r \in \{-1, 1\}$ can be written with the use of the operator $\hat{\wp}$ whose activity on $|0\rangle$ is such that

$$r = \langle 0 | sign(\hat{\wp}) | 0 \rangle \tag{12}$$

Hence, the arithmetic algebraic form of the operator, $\hat{\wp}$, is provided by Eq. (9), the vector-matrix form like in (12).

If we write e.g. for $A_a(\lambda, \mu)$ the following

$$A_a(\lambda, \mu) = \mu \langle 0 | sign(\kappa + S_3 e^{\zeta_A}) | 0 \rangle \tag{13}$$

and

$$A_b(\lambda, \mu) = S_0 \langle 0 | sign(\kappa + S_4 e^{\zeta_B}) | 0 \rangle \tag{14}$$

Here, $S_i \in \{-1, 1\}$, $(i = 3, 4)$ and, $S_3 + S_4 = 0$. I.e. the source sends e.g., $S_3 = -1$ to A and $S_4 = +1$ to B. With, $\kappa = \exp[\frac{1}{2} e^{-(\lambda - \log(4)/2)}]$.

Because of, $F_{a,b}(\lambda) = \sum_{\mu \in \{-1,1\}} F_{a,b}(\lambda, \mu) \delta_{\mu, r}$, and $F_{a,b}(\lambda, \mu) = A_a(\lambda, \mu) A_b(\lambda, \mu)$, it follows

$$F_{a,b}(\lambda) = S_0 \langle 0 | sign(\hat{\wp}) | 0 \rangle \langle 0 | sign(\kappa + S_3 e^{\zeta_A}) | 0 \rangle \langle 0 | sign(\kappa + S_4 e^{\zeta_B}) | 0 \rangle \tag{15}$$

From, $|0\rangle\langle 0| = \hat{1}$, it is possible to re-write the previous equation into

$$F_{a,b}(\lambda) = S_0 \langle 0 | sign(\hat{\wp}) sign(\kappa + S_3 e^{\zeta_A}) sign(\kappa + S_4 e^{\zeta_B}) | 0 \rangle \tag{16}$$

Making use of, $sign(xy) = sign(x) sign(y)$, the previous can be re-written as

$$F_{a,b}(\lambda) = S_0 \langle 0 | sign[\hat{\wp}(\kappa + S_3 e^{\zeta_A})(\kappa + S_4 e^{\zeta_B})] | 0 \rangle \tag{17}$$

First we may deduce from the algebraic form of operator, $\hat{\wp}$, that

$$\hat{\wp} \kappa^2 = -\log \frac{1}{2} \sum_{S_1 \in \{-1,1\}} \frac{1}{2} \sum_{S_2 \in \{-1,1\}} \log^{-1} 2 \log \log \exp[e^{-(\lambda - \log(4))/2}]$$
$$= -\log e^{-(\lambda - \log(4))} = \lambda - \log(4)$$

In addition,

$$\hat{\wp} \kappa e^{\zeta_A} = \hat{\wp} \exp[\tfrac{1}{2} e^{-(\lambda - \log(4))/2} + \zeta_A]$$
$$= -\log \frac{1}{2} \sum_{S_1 \in \{-1,1\}} \frac{1}{2} \sum_{S_2 \in \{-1,1\}} \log^{-1} \log[\tfrac{1}{2} e^{-(\lambda - \log(4))/2} + \zeta_A]^2$$
$$= -\log \frac{1}{2} \sum_{S_1 \in \{-1,1\}} \frac{1}{2} \sum_{S_2 \in \{-1,1\}} [\zeta^2_A + \zeta_A e^{-(\lambda - \log(4))/2} + \tfrac{1}{4} e^{-(\lambda - \log(4))}]$$
$$= -\log(1 + \tfrac{1}{4} e^{-(\lambda - \log(4))})$$

Hence, $\hat{\wp}\kappa e^{\zeta_A} = \hat{\wp}\kappa e^{\zeta_B}$ and because, $S_3 + S_4 = 0$, the two terms cancel. Because of the fact that, $S_3 S_4 = -1$, equation (11) arises.

## III Correlation

Having sufficiently established the expression in Eq. (11) we may derive the following form of $I_{a,b}$

$$I_{a,b} = S_0 \int_0^{+\infty} 2(1-e^{-\lambda})\delta(\lambda - \hat{\wp}e^{\zeta_A}e^{\zeta_B} - \log(4))d\lambda \tag{18}$$

because, $(d\,sign(x)/dx) = 2\delta(x)$. From the previous equation (11) we may note that, because, $(1-e^{-\lambda}) \geq 0,\ \forall_{\lambda \in [0,\infty)}$ and, $\delta(x) \geq 0,\ \forall_{x \in \mathbb{R}}$, that, $sign(I_{a,b}) = S_0 \in \{-1,1\}$.

If we subsequently compute the result of the integral in equation (11) we see that

$$\hat{\wp}e^{\zeta_A}e^{\zeta_B} = -\log\frac{1}{2}\sum_{S_1 \in \{-1,1\}}\frac{1}{2}\sum_{S_2 \in \{-1,1\}}\log^{-1}2\log\log(e^{\zeta_A+\zeta_B})$$

Hence

$$\hat{\wp}e^{\zeta_A}e^{\zeta_B} = -\log\frac{1}{2}\sum_{S_1 \in \{-1,1\}}\frac{1}{2}\sum_{S_2 \in \{-1,1\}}\log^{-1}2\log(\zeta_A+\zeta_B) =$$

$$-\log\frac{1}{2}\sum_{S_1 \in \{-1,1\}}\frac{1}{2}\sum_{S_2 \in \{-1,1\}}(\zeta_A+\zeta_B)^2 = -\log[2(1+(a.b)S_0)]$$

From eq. (7) it then follows that

$$C_{a,b} = sign(I_{a,b})(1-(a.b)S_0 - 1) = -S_0 sign(I_{a,b})(a.b) = -(a.b) \tag{19}$$

which can be recognized as the quantum correlation.

## IV Discussion

In the previous paper the quantum correlation is derived from Bell's expression for Local Hidden Causal variables such as envisaged by Einstein.

The reader should note that there is little reason to question the employed exponential density and the Kronecker discrete density as classical densities. Hence, we can concentrate on the measurement functions. In the first place because the present derivation is simple and straightforward, the question 'what is wrong with Bell's inequality' is that this inequality should forbid to have,

$$F_{a,b}(\lambda) = S_0\, sign(\lambda - \hat{\wp} e^{\zeta_A} e^{\zeta_B} - \log(4))$$

This is an interesting claim. Its reasoning would go something like,.. 'because of the truth of Bell's inequality –based on e.g. its simplicity- it is not possible to write this'. However, arithmetically this is possible. This implies that the expression could be a breach of the locality –admitting then at the same time that the inequality can be violated with non-local LHV models thereby weakening the 'what is wrong argument' to the issue of locality v.s. non-locality leaving aside the implied mathematical error intended by the opponent. In simple words, the opponent would then resort to … 'breaches of the inequality are not mathematical errors when a non-local LHV model is employed'. This brings us to the question if the expression is local or non-local. Before embarking on this part of the discussion, it is necessary to note that explicit forms of A and B measurement functions were given in section II. Perhaps that critical readers still will be dissatisfied because here it is still not quite possible to point at a technical error, but the formulation admittedly has the drawback of the unknown[2]

$$r = \langle 0 | sign(\hat{\wp}) | 0 \rangle$$

In order to accommodate those readers I refer to a model in reference [5] and again stress that in reference [7] classical probability also did something that, no doubt, will be considered impossible by this type of reader.

Let us try to answer the locality vs non-locality question as follows.

A product of two a, b parameter-independent functions, like e.g.

$$A_a(\lambda, \mu) A_b(\lambda, \mu)$$

is considered local because otherwise Bell's correlation would not be local from the beginning. Hence, the product, $e^{\zeta_A} e^{\zeta_B}$ passes the test of locality too. Then, any function on $e^{\zeta_A} e^{\zeta_B}$ will not represent a breach of locality either, especially when the parameters of the function –i.e. the 'stuff' where the function is made of- is independent of the parameter vectors. Note that, of course in the A-wing, the parameter vector is chosen independently of the choice in the B-wing and vice-versa. Hence, the expression $F_{a,b}(\lambda) = S_0\, sign(\lambda - \hat{\wp} e^{\zeta_A} e^{\zeta_B} - \log(4))$ does not contain non-local elements because, firstly, the independent choice of parameter vectors in the respective wings is absolutely maintained. Moreover, the source only 'blindly' sends hidden parameters $S_0, S_1, S_2$ to the A-wing and the B-wing. Those parameters are employed in the 'function' that transforms $e^{\zeta_A} e^{\zeta_B}$.

Suppose, furthermore, we inspect a computer network model of Gill [3].

---

[2] Of course random drawing of numbers in {-1,1} could help to overcome this lack of knowledge. Best is to make a computer model based on ref [5].

Let us study the Clauser, Horne, Shimony and Holt variant of the Bell inequalities. We have

$$D = C_{1_A,1_B} - C_{1_A,2_B} - C_{2_A,1_B} - C_{2_A,2_B}$$

According to a similar argument as for inequality (2), $D$, is restricted to, $|D| \leq 2$. Here, the observer A in the A-wing may randomly select from unitary, {a,b} and B from the set {c,d} and, {a,b}≠{c,d}. It should be noted that the previous inequality can be violated with only a limited set of choices for {a,b}≠{c,d}. In the computer experiment, we restrict the attention to a two-dimensional sub-space and take

$$a = 1_A = (1,0), \qquad b = 2_A = (0,1)$$

$$c = 1_B = (1/\sqrt{2}, -1/\sqrt{2}), \qquad d = 2_B = (-1/\sqrt{2}, -1/\sqrt{2})$$

Note that e.g. for the pair $(1_A, 1_B)$, the quantum correlation is equal to $C_{1_A,1_B} = -\frac{1}{\sqrt{2}}$.

For our expression $C_{a,b} = sign(I_{a,b})(|I_{a,b}| - 1)$ we may see the following

$$sign(I_{1_A,1_B}) = -1, \ sign(I_{a,b}) = 1, \ for \ all \ (a,b) \in \{(1_A, 2_B), (2_A, 1_B), (2_A, 2_B)\} \equiv \Re$$

together with

$$|I_{a,b}| - 1 = \frac{1}{\sqrt{2}}, \ \forall_{(a,b) \in \Re \cup \{(1_A,1_B)\}} \tag{20}$$

This then gives, $D = \left|-\frac{1}{\sqrt{2}} - \frac{1}{\sqrt{2}} - \frac{1}{\sqrt{2}} - \frac{1}{\sqrt{2}}\right| = 2\sqrt{2} > 2$. This shows that with the use of the previous formalism at least one of the (computer simulation) experiments (Gill [3]) can be violated with a local hidden causality model. This validates the previously given formalism because a simple violation can be obtained from the formalism. This one-time violation using the given set of settings for the measuring instruments, already is sufficient enough when the opponents are unable to show non-locality in the formalism. The truth of the auxiliary argument in section II has no dealing with this. If critical readers contest this result then they cannot do that unwarranted. It should be demonstrated that either, locality is violated, or, mathematical errors were made, or, that equation (20) is, out of internal mathematical reasons, impossible –naturally without referring to the truth of CHSH or using section II arguments-, or, revert to the authoritarian version of religion.

Concerning the other important argument against local hidden causality, i.e. the value definiteness and non-contextuality of the Kochen and Specker argument, one can employ the possibility that upon measurement the quantum variables are 'filled' with values from a conglomerate of deterministic local hidden variables and that there is no value definiteness in the quantum variables but that there is in the local hidden causality that 'feeds' them. In this way LHVs may avoid the value-definiteness of the Kochen and Specker theorem leading to a possible conflict between LHV models and quantum predictions.

In general we have demonstrated that along the previous lines of reasoning one can obtain the quantum correlation from a local hidden variables expression of the correlation. The conclusion must be that Bell's mathematics is not that unqualified universally valid as is usually assumed. The non-localists cannot hide forever behind the simplicity of Bell's argument, turn a blind eye to its shortcomings and uncritically trust on the operations leading to the inequality. The density is unquestionably classic Kolmogorovian and hence the advanced criticism on Bell's mathematics cannot be rejected along the lines of probability theory. The here employed formalism remains within the boundaries of locality. The present author would like to make the reader aware of other counter-models (Geurdes [5]).

Locality has earned its place among the possible explanations of quantum correlation. Moreover, the author would like to point at the fact that, mathematically, Dirac's quantum relativistic equation appears embedded in Maxwell's classical electromagnetic field equations [6]. This at its least implies a less straightforward theoretical relation between the classical and the quantum domain.